\documentclass[amsmath,amssymb,aps,10pt,prd,twocolumn,letterpaper,nofootinbib,balancelastpage,notitlepage,superscriptaddress,floatfix,preprintnumbers]{revtex4-2}
\usepackage{graphicx}	
\usepackage{ragged2e}	
\usepackage{bm}		    
\usepackage[sort&compress]{natbib}	 	
	\setcitestyle{square,numbers,comma}	
\usepackage[colorlinks=true,urlcolor=blue,linkcolor=black,citecolor=red]{hyperref}	
\newcommand{\figref}[2][]{Fig{#1}.~\ref{fig:#2}}		
\newcommand{\secref}[2][]{Sec{#1}.~\ref{sec:#2}}		
\renewcommand{\eqref}[2][]{Eq{#1}.~(\ref{eq:#2})}		
\newcommand{\eqrefRange}[2]{Eqs.~(\ref{eq:#1})--(\ref{eq:#2})}		
\newcommand{\citeR}[2][]{Ref{#1}.~\cite{#2}}			
\newcommand{\orcid}[1]{\href{https://orcid.org/#1}{\,\includegraphics[width=8px]{ORCID.png}}}
\newcommand{\N}{\bm{\hat n}}
\newcommand{\Z}{\bm{\hat z}}

\begin{document}

\title{Tunneling away the relic neutrino asymmetry}
\date{\today}
\author{Saarik Kalia}
\email{kalias@umn.edu}
\affiliation{School of Physics \& Astronomy, University of Minnesota, Minneapolis, MN 55455, USA}

\begin{abstract}
The Earth acts as a matter potential for relic neutrinos which modifies their index of refraction from vacuum by $\delta\sim10^{-8}$.  It has been argued that the refractive effects from this potential should lead to a large $\mathcal O(\sqrt\delta)$ neutrino-antineutrino asymmetry at the surface of the Earth.  This result was computed by treating the Earth as flat.  In this work, we revisit this calculation in the context of a perfectly spherical Earth.  We demonstrate, both numerically and through analytic arguments, that the flat-Earth result is only recovered under the condition $\delta^{3/2}kR\gg1$, where $k$ is the typical momentum of the relic neutrinos and $R$ is the radius of the Earth.  This condition is required to prevent antineutrinos from tunneling into classically inaccessible trajectories below the Earth's surface and washing away the large asymmetry.  As the physical parameters of the Earth do not satisfy this condition, we find that the asymmetry at the surface should only be $\mathcal O(\delta)$.  While the asphericity of the Earth may serve as a loophole to our conclusions, we argue that it is still difficult to generate a large asymmetry even in the presence of local terrain.
\end{abstract}

\maketitle

\section{Introduction}
\label{sec:introduction}

Standard cosmology predicts the existence of a cosmic neutrino background (C$\nu$B) of relic neutrinos produced in the early universe, following a Fermi-Dirac distribution with temperature $T_\nu=1.7\times10^{-4}\,\mathrm{eV}$~\cite{kolb,baumann}.  While the detection of the C$\nu$B would lead to profound insights about the fundamental properties of neutrinos and early universe cosmology~\cite{yanagisawa,scott}, its direct detection via scattering/absorption is difficult because neutrino cross sections scale as $\mathcal O(G_F^2)$~\cite{weinberg,ptolemy}.  Another proposal was put forth by Stodolsky to look for electron/nuclear spin precession due to the C$\nu$B~\cite{stodolsky}.  While this effect is $\mathcal O(G_F)$, it is proportional to the neutrino-antineutrino asymmetry, which in vacuum is expected to be comparable to the observed baryon asymmetry of $\sim10^{-9}$.

In \citeR{arvanitaki}, it was proposed that the neutrino-antineutrino asymmetry may be significantly enhanced near the surface of the Earth.  This enhancement occurs because neutrinos/antineutrinos experience a potential
\begin{equation}
    |U|\sim10^{-14}\,\mathrm{eV}\cdot\left(\frac{n_\mathrm{matter}}{10^{22}\,\mathrm{cm}^{-3}}\right)
\end{equation}
in the presence of matter.  For electron neutrinos and muon/tau antineutrinos, this potential is positive, while for electron antineutrinos and muon/tau neutrinos, it is negative.  This leads to an index of refraction for neutrinos/antineutrinos inside the Earth which differs from vacuum by
\begin{equation}
    \delta\equiv-\frac{m_\nu U}{k^2}\sim\pm10^{-8}\cdot\left(\frac{n_\mathrm{matter}}{10^{22}\,\mathrm{cm}^{-3}}\right)\cdot\left(\frac{m_\nu}{0.1\,\mathrm{eV}}\right),
\end{equation}
where $k\sim T_\nu$ is the momentum of the neutrino and $m_\nu$ is the mass of the relevant neutrino species.  Due to the different refractive indices for neutrinos vs. antineutrinos, it is argued in \citeR{arvanitaki} that the Earth induces a $\mathcal O(\sqrt\delta)\sim10^{-4}$ fractional asymmetry near its surface.  Importantly, the calculation in \citeR{arvanitaki} treated the Earth as flat.

In this work, we recompute this asymmetry in the context of a perfectly spherical Earth.  While similar computations have been performed in \citeR[s]{huang,gruzinov,vantilburg2024wake}, the principle purpose of this work is to show that the large asymmetry observed in the flat-Earth calculation is only reproduced under the condition
\begin{equation}\label{eq:condition}
    \delta^{3/2}kR\gg1,
\end{equation}
where $R$ is the radius of the Earth.  The physical parameters of the Earth have $\delta^{3/2}kR\sim0.01$, and so do not satisfy \eqref{condition}.  In this case, we show that a much smaller $\mathcal O(\delta)$ fractional asymmetry should be expected at the surface of the Earth.  Our calculation treats the Earth as a perfect sphere, and so the presence of local terrain may provide a loophole to the above condition.  In our concluding discussion, however, we comment that local terrain should typically make the condition more stringent rather than less.

This work is organized as follows.  In \secref{heuristic}, we give a heuristic understanding, in terms of ray optics, of how the condition in \eqref{condition} arises.  In \secref{spherical}, we perform a precise calculation of the neutrino-antineutrino asymmetry, under the assumption of a perfectly spherical Earth of uniform density.  As in \citeR{arvanitaki}, for this calculation, we consider only electron neutrinos/antineutrinos and treat the C$\nu$B as monochromatic and isotropic.  In \secref{discussion}, we comment on how the asphericity of the Earth affects the interpretation of our results and contrast our work with \citeR[s]{huang,gruzinov,vantilburg2024wake}.  We make all the code used in this
work publicly available on Github~\cite{github}.

\section{Heuristic argument}
\label{sec:heuristic}

\begin{figure*}[t]
\includegraphics[width=\textwidth]{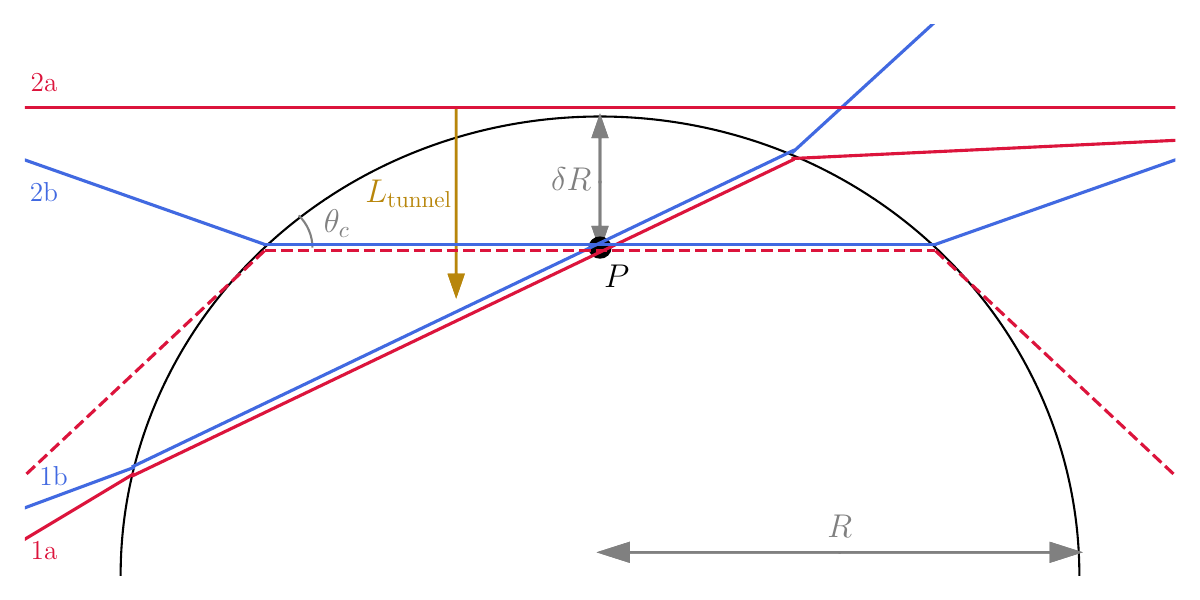}
\caption{\label{fig:earthfig}%
    Diagram of the heuristic argument presented in \secref{heuristic} (not to scale).  The asymmetry at a point $P$ receives contributions from all rays passing through it.  Blue rays represent neutrino paths, while red rays represent antineutrino paths.  Rays that arrive at large angles, such as those labeled 1a and 1b, traverse the interior of the Earth in pairs, contributing a small asymmetry at $P$.  However, neutrino rays that form angles $\theta<\theta_c$ with the surface of the Earth, such as ray 2b, have no corresponding antineutrino ray, as antineutrinos cannot classically access this path in the interior of the Earth.  This can lead to a large asymmetry for all points $P$ within $\delta R$ of the surface.  Quantum mechanically, however, the ray 2a can tunnel to this classically inaccessible trajectory, thereby washing out the asymmetry.  This can occur when its characteristic tunneling range $L_\mathrm{tunnel}\sim(kR)^{1/3}/k$ [see \eqref{tunnel}] exceeds the required distance $\delta R$ that it must tunnel, or in other words, when $\delta^{3/2}kR\ll1$.  See text for more details.
    }
\end{figure*}

We begin by presenting a heuristic understanding of the full calculation presented in \secref{spherical}.  The relevant features of the argument are depicted in \figref{earthfig}.  In order to determine the asymmetry at the surface of the Earth, let us consider a point $P$ located inside the Earth but close to the surface.  If the asymmetry at this point is large, then by continuity, the asymmetry at the surface will be as well.  The neutrino and antineutrino densities at $P$ receive contributions from all of the rays passing through $P$.%
\footnote{For illustrative purposes, we phrase the argument in this section in terms of individual rays, but more properly, the detailed calculation presented in the next section will deal with angular modes.  As all rays with the same impact parameter (with respect to the center of the Earth) possess the same angular momentum, this argument should be understood in a context where all rays with the same impact parameter are considered equivalent.  For instance, the relative orientations of rays 2a and 2b in \figref{earthfig} are not particularly important.  What is relevant are their impact parameters.}
For every antineutrino ray, there is a corresponding neutrino ray which travels the same path through the interior of the Earth; e.g., see the rays labeled 1a and 1b in \figref{earthfig}.  The contributions from these rays cancel, leading to no significant asymmetry.%
\footnote{In fact, due to their different incident angles outside the Earth, these rays possess slightly different normalizations which results in an $\mathcal O(\delta)$ asymmetry.}

Consider instead a neutrino ray which traverses a chord very close to the surface of the Earth; e.g., ray 2b in \figref{earthfig}.  For such rays there exists no corresponding antineutrino ray.  This is because such an antineutrino ray attempting to emerge from the interior of the Earth would experience total reflection, as it exits from a medium of higher index of refraction to one of a lower index.  Conversely, this implies that classically no antineutrino ray originating from outside the Earth can reach this trajectory in the interior.  These unpaired neutrino rays are the source of the large asymmetry described in \citeR{arvanitaki}.  These rays are unpaired so long as they form an angle $\theta<\theta_c\equiv\sqrt{2|\delta|}$ with the surface of the Earth (from the interior; see \figref{earthfig}).  All points $P$ which are within a distance $\theta_c^2R/2=|\delta |R$ from the surface will receive a contribution from such rays, and so will exhibit a large asymmetry.  By continuity, points just above the surface will exhibit this asymmetry as well.

The above argument only holds in the context of classical ray optics, where neutrino/antineutrinos travel on fixed paths determined by refraction.  When quantum wave effects are included, antineutrinos have the ability to tunnel from a ray outside the Earth onto this classically forbidden trajectory inside the Earth; e.g., ray 2a in \figref{earthfig} may tunnel inside the Earth to cancel the contribution from ray 2b.  If this tunneling occurs efficiently, the large asymmetry will be washed out.  Conservation of energy and angular momentum relate the parameters of the external and internal trajectories.  Suppose the external trajectory has a velocity $v_1$ and impact parameter $r_1$, while the internal trajectory has a velocity $v_2$ and impact parameter $r_2$.  Then, we have
\begin{align}
    \frac{m_\nu v_1^2}2&=\frac{m_\nu v_2^2}2-U\\
    m_\nu v_1r_1&=m_\nu v_2r_2,
\end{align}
which implies
\begin{equation}
    r_2=\frac{r_1}{\sqrt{1+\frac{2U}{m_\nu v_1^2}}}\approx(1-\delta)r_1.
\end{equation}
In other words, the ray must tunnel a distance roughly $\delta R$ from its trajectory outside the Earth to its trajectory inside the Earth.  As we will see in \secref{spherical}, the typical distance that a ray can efficiently tunnel is $L_\mathrm{tunnel}\sim(kR)^{1/3}/k$.  Therefore, when
\begin{equation}
    L_\mathrm{tunnel}\ll\delta R\implies\delta^{3/2}kR\gg1,
\end{equation}
the antineutrino rays outside the Earth will be unable to tunnel in, and so the large asymmetry will persist.

\section{Spherical calculation}
\label{sec:spherical}

Now we make the arguments of \secref{heuristic} more precise by performing a calculation of the neutrino-antineutrino asymmetry near the Earth's surface, under the assumption of a perfectly spherical Earth.  The formalism introduced in this section will be similar to the formalisms applied in \citeR[s]{huang,gruzinov,vantilburg2024wake}, but we will focus on determining the conditions under which the asymmetry exists.  As mentioned in \secref{introduction}, we will assume that the Earth is a perfect sphere of uniform density, and that the C$\nu$B is monochromatic (with momentum $k$) but isotropic.%
\footnote{We note that the C$\nu$B is not in fact isotropic in the rest frame of the Earth, due to their relative motion.  As the thermal velocity of relic neutrinos $v_\mathrm{th}\sim T_\nu/m_\nu\sim10^{-3}\cdot(m_\nu/0.1\,\mathrm{eV})$ may be comparable to this relative motion, this anisotropy may be significant, depending on the mass of the relevant neutrino species.  In this work, we will neglect this potential anisotropy.}
We will also compute the time-averaged asymmetry, in which case, it suffices to solve for the steady-state neutrino/antineutrino wavefunctions.  In order to determine these wavefunctions at a given point for an isotropic background, it is sufficient to compute the solution for a single incoming plane wave and average this solution over a sphere of constant radius.  More precisely, given an incoming plane wave from direction $\N$, let $\psi_{\N}(r,\Omega)$ denote the resulting wavefunction in the vicinity of the Earth.  By symmetry, it is simple to see that $\psi_{\N}(r,\Z)=\psi_{\Z}(r,\N')$, where $\N'$ is $\N$ rotated by $\pi$ around the $z$-axis.  The total neutrino/antineutrino density is given by the average of $|\psi_{\N}|^2$ over all $\N$, so we can write
\begin{align}
    \frac{n_{\nu,\bar\nu}(r)}{n_0}&=\frac1{4\pi}\int d\N\,|\psi_{\N}(r,\Z)|^2\\
    &=\frac1{4\pi}\int d\Omega\,|\psi_{\Z}(r,\Omega)|^2,
\label{eq:density}\end{align}
where $n_{\nu,\bar\nu}$ represents the neutrino/antineutrino density, and $n_0$ denotes their density in the absence of the Earth.  We see then that it is sufficient to compute only the solution $\psi_{\Z}$ for a plane wave incoming from the $\Z$-direction.  Henceforth we will simply denote this solution by $\psi$.

Let us decompose this wavefunction into angular modes as
\begin{equation}
    \psi(r,\Omega)=\sum_{\ell=0}^\infty\psi_\ell(r)Y_{\ell0}(\Omega),
\end{equation}
where we only require $m=0$ modes by azimuthal symmetry.  The Schrodinger equation for the radial part $\psi_\ell$ of the wavefunction becomes
\begin{equation}\label{eq:schrodinger}
    \partial_r^2\psi_\ell+\frac2r\partial_r\psi_\ell+\left(k^2-2mU\cdot\Theta(R-r)-\frac{\ell(\ell+1)}{r^2}\right)\psi_\ell=0,
\end{equation}
The solution to \eqref{schrodinger} is simply a combination of spherical Bessel functions, with momentum $k$ for $r>R$ and momentum
\begin{equation}
    k'\equiv\sqrt{k^2-2mU}=k\sqrt{1+2\delta}
\end{equation}
for $r<R$.  The corresponding boundary conditions for \eqref{schrodinger} should be that the only incoming modes are that of the incoming plane wave, and that the wavefunction is regular at the origin.  An incoming plane wave from the $\Z$-direction can be decomposed in terms of spherical harmonics as
\begin{equation}
    e^{ikz}=\sum_{\ell=0}^\infty i^\ell\sqrt{4\pi(2\ell+1)}j_\ell(kr)Y_{\ell0}(\Omega),
\end{equation}
where $j_\ell$ is the spherical Bessel function of the first kind.  The full solution for the wavefunction in and around the Earth should therefore be
\begin{align}
    \psi_\ell(r)&=\left\{\begin{array}{lc}A_\ell j_\ell(kr)+B_\ell h_\ell^{(1)}(kr),&r>R\\C_\ell j_\ell(k'r),&r<R,\end{array}\right.\\
    A_\ell&\equiv i^\ell\sqrt{4\pi(2\ell+1)},
\end{align}
where the spherical Hankel function $h_\ell^{(1)}$ of the first kind corresponds to an outgoing wave.

Continuity of $\psi_\ell$ and its gradient enforce
\begin{align}\label{eq:boundary1}
    A_\ell j_\ell(kR)+B_\ell h^{(1)}_\ell(kR)&=C_\ell j_\ell(k'R)\\
    kA_\ell j'_\ell(kR)+kB_\ell h^{(1)\prime}_\ell(kR)&=k'C_\ell j'_\ell(k'R).
\label{eq:boundary2}\end{align}
These can be solved numerically for sufficiently large $\ell$ to determine the full solution $\psi$.  From \eqref{density}, we see that the number density is then given by
\begin{align}
    \frac{n_{\nu,\bar\nu}(r)}{n_0}&=\frac1{4\pi}\sum_{\ell=0}^\infty|\psi_\ell(r)|^2\\
    &=\left\{\begin{array}{lc}\frac1{4\pi}\sum_\ell\left|A_\ell j_\ell(kr)+B_\ell h_\ell^{(1)}(kr)\right|^2,&r>R\\\frac1{4\pi}\sum_\ell\left|C_\ell j_\ell(k'r)\right|^2,&r<R,\end{array}\right.
\label{eq:density2}\end{align}
Using $\delta<0$ to compute the above solution will give the neutrino density $n_\nu$, while using $\delta>0$ will give the anti-neutrino density $n_{\bar\nu}$.  By taking the difference of \eqref{density2} between these two cases, we can compute the fractional neutrino asymmetry $\Delta=(n_\nu-n_{\bar\nu})/n_0$.

\begin{figure}[t]
\includegraphics[width=0.99\columnwidth]{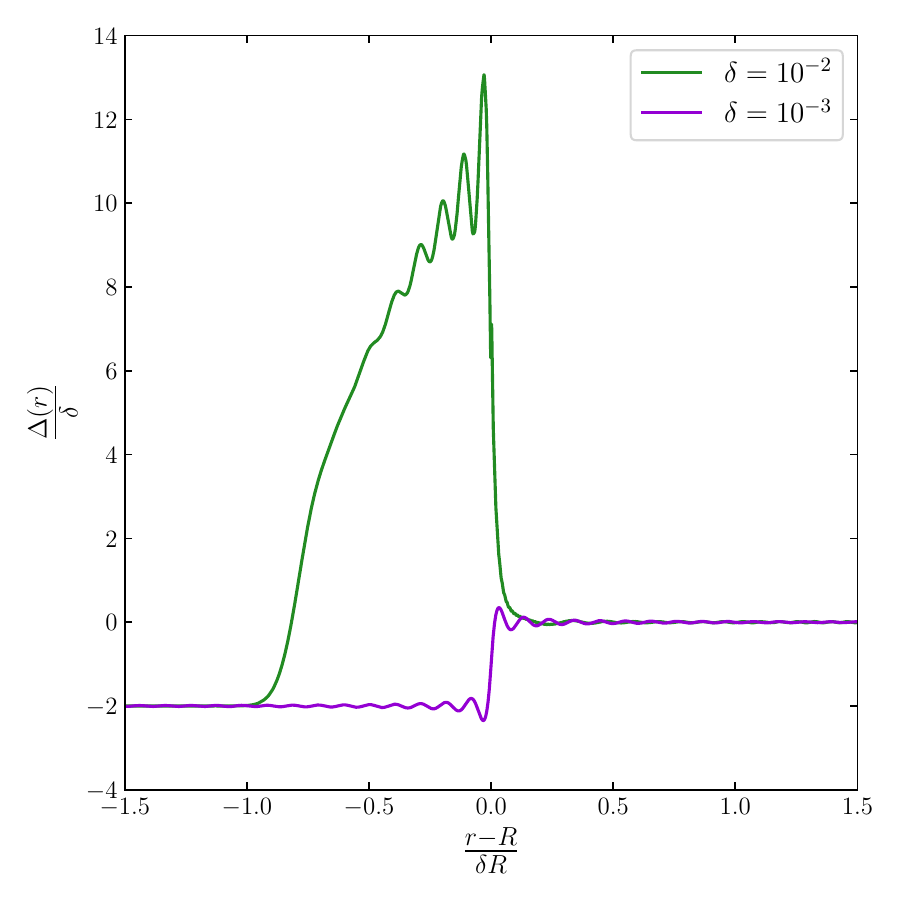}
\caption{\label{fig:profiles}%
    Plot of the neutrino-antineutrino asymmetry $\Delta(r)$ [normalized by $\delta$] near the surface of the Earth, for $kR=3\times10^4$ and two different values of $\delta$.  [To compute these profiles, we sum up to $\ell=3.2\times10^4$ in \eqref{density2}.]  In both cases, we see $\Delta=-2\delta$ for $r<(1-\delta)R$ and $\Delta=0$ well above the surface of the Earth.  In the $\delta=10^{-2}$ case, however, we see that the asymmetry becomes large and positive for $(1-\delta)R<r\leq R$.
    }
\end{figure}

\begin{figure}[t]
\includegraphics[width=0.99\columnwidth]{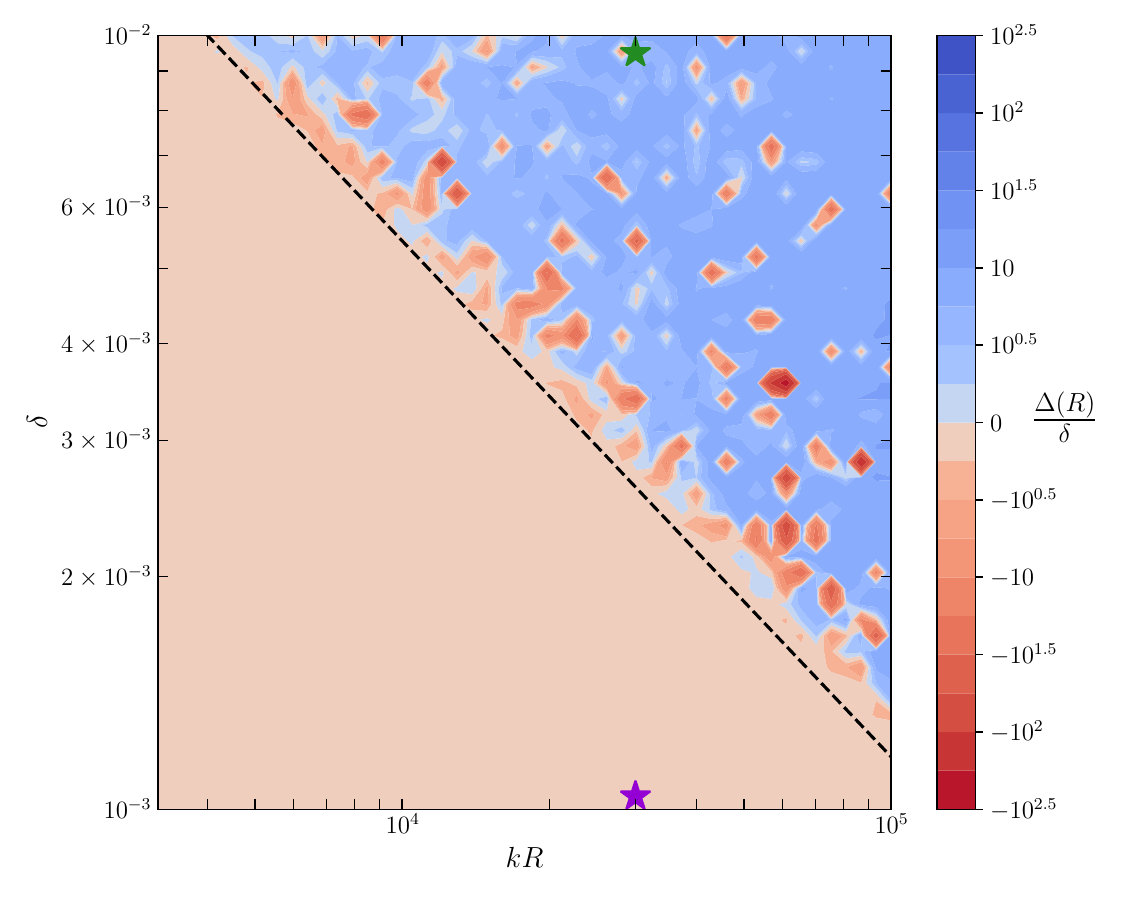}
\caption{\label{fig:contour}%
    Contour plot of the asymmetry at the surface of the Earth $\Delta(r=R)$ [again normalized by $\delta$] for various values of $kR$ and $\delta$.  The dashed black line denotes $\delta^{3/2}kR=4$, and the green and purple stars correspond to the parameter values used in \figref{profiles}.  Note that large asymmetries are only obtained above the line.  Below the line, $\Delta(R)=-\delta$ for all parameter values.  The physical parameters for the Earth have $\delta^{3/2}kR\sim0.01$ and so fall well below the black line.
    }
\end{figure}

The sums in \eqref{density2} converge only after $\ell\gtrsim kR$, so evaluating them for the physical parameters of the Earth ($kR\sim9\times10^9$) is computationally challenging.  \citeR{huang} claims to have achieved this with the use of unspecified asymptotic Bessel function expansions.  In this work, we instead opt to understand the relevant behavior by evaluating the solution for smaller parameter values.  In \figref{profiles}, we compute the asymmetry $\Delta(r)$ as a function of radius, for $kR=3\times10^4$ and two choices of $\delta=10^{-2}$ and $\delta=10^{-3}$.  A number of important features can be seen even with these parameter values.  In both cases, the asymmetry begins at $\Delta=-2\delta$ deep within the Earth and eventually reaches $\Delta=0$ outside the Earth.  In the case of small $\delta$, the profile transitions directly between these regimes, leading to an asymmetry of $\Delta=-\delta$ at the surface $r=R$.%
\footnote{The profile exhibits oscillations near the surface of the Earth, but these will be smoothed out after averaging over $k$.}
In the case of large $\delta$, however, the asymmetry becomes large and positive in the region $(1-\delta)R<r\leq R$, leading to an asymmetry larger than $\mathcal O(\delta)$ at $r=R$.%
\footnote{The exact shape and size of this large asymmetry can depend on $kR$ and $\delta$.  In \citeR{arvanitaki}, it is argued that this asymmetry is $\mathcal O(\sqrt\delta)$ and extends for a distance $\lambda_c\equiv2\pi/\theta_ck$ above the surface of the Earth.  We find similar results in our numerical computations, but as the purpose of this work is to understand when this large asymmetry exists, we choose not to focus on how large it is or how far it extends.}
In \figref{contour}, we show the asymmetry at the Earth's surface $\Delta(r=R)$ for various values of $kR$ and $\delta$.  This plot clearly demonstrates that a large asymmetry at the surface is only achieved when $\delta^{3/2}kR\gg1$.  Below, we justify why this is the case.

First, let us derive the $\Delta=-2\delta$ asymmetry for $r<(1-\delta)R$.  It is straightforward to solve \eqref[s]{boundary1} and (\ref{eq:boundary2}) to find
\begin{equation}\label{eq:Cl1}
    C_\ell=\frac{iA_\ell k^{-2}R^{-2}}{h^{(1)\prime}_\ell(kR)j_\ell(k'R)-\sqrt{1+2\delta}\cdot h^{(1)}_\ell(kR)j'_\ell(k'R)}.
\end{equation}
Now the spherical Bessel and Hankel functions satisfy
\begin{gather}
    \left|h_\ell^{(1)}(x)\right|^2\approx\left|h_\ell^{(1)\prime}(x)\right|^2\approx j_\ell(x)^2+j'_\ell(x)^2\approx\frac1{x^2},\\
    \langle j_\ell(x)^2\rangle\approx\frac1{2x^2},
\end{gather}
for $x>\ell$, where $\langle\cdot\rangle$ indicates an average over $\ell$.  With these approximations, \eqref{Cl1} becomes
\begin{equation}\label{eq:Cl2}
    \langle|C_\ell|^2\rangle\approx4\pi(2\ell+1)(1+\delta),
\end{equation}
for $\ell<kR$ and $\delta\ll1$.  Plugging \eqref{Cl2} into \eqref{density2} gives
\begin{equation}\label{eq:inside}
    \frac{n_{\nu,\bar\nu}(r)}{n_0}\approx(1+\delta)\sum_{\ell=0}^\infty(2\ell+1)j_\ell(k'r)^2=1+\delta.
\end{equation}
Subtracting the neutrino and antineutrino cases gives the asymmetry $\Delta=-2\delta$.  The sum in \eqref{inside} is dominated by $\ell<k'r$, because as we will see momentarily, a mode with angular momentum $\ell$ does not penetrate to radii smaller than $\ell/k'$.  Therefore, in order to apply the approximation in \eqref{Cl2}, we require $k'r<kR$.  In other words, this asymmetry only holds for $r<(1-\delta)R$.

\begin{figure}[t]
\includegraphics[width=0.99\columnwidth]{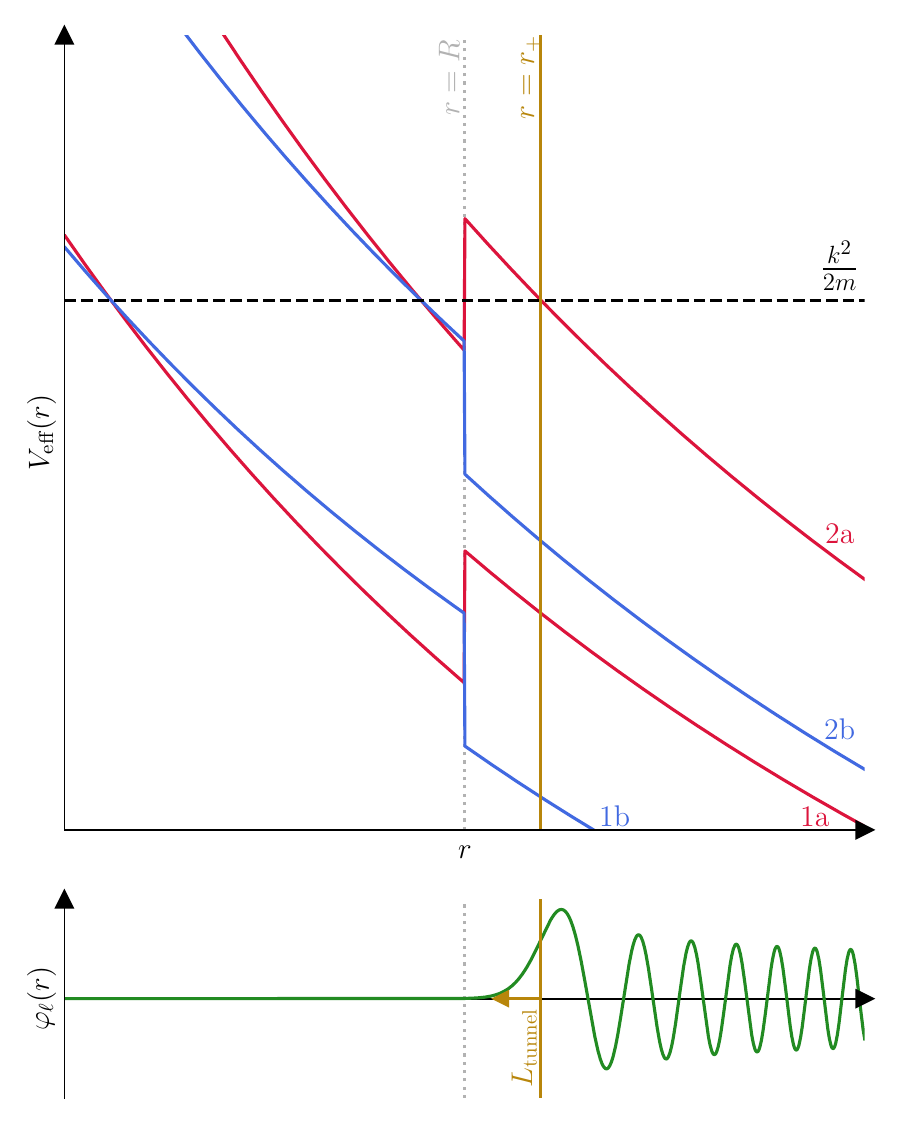}
\caption{\label{fig:potentials}%
    Schematic diagram of several effective potentials $V_\mathrm{eff}$ near the surface of the Earth.  The potentials for neutrino modes are shown in blue, while the potentials for antineutrino modes are shown in red, and the potentials are labeled to correspond schematically to the rays shown in \figref{earthfig}.  The dashed black line denotes the energy of the incoming wave.  The point where a given potential meets the dashed line is its classical ``turning point" $r_0$, and the potentials have been paired so that they share the same turning point inside the Earth.  Potentials 1a and 1b have low (but slightly different) values of $\ell$, and so their turning point lies deep within the Earth.  Neutrino potential 2b, on the other hand, has a turning point $(1-\delta)R<r_0<R$.  The corresponding antineutrino potential 2a exhibits a second turning point $r_+$ outside the Earth, so that an incident wave will not be able to penetrate into the Earth classically.  Quantum mechanically, however, the wave may tunnel through the barrier of potential 2a to reach the interior of the Earth.  In the lower plot, we show an incoming wave incident on potential 2a.  The characteristic length scale over which it decays once entering the barrier is $L_\mathrm{tunnel}$ [see \eqrefRange{WKB1}{tunnel}].  In the case shown here, $L_\mathrm{tunnel}<r_+-R$, so that the wave cannot tunnel into the Earth.  For the physical parameters of the Earth, however, $L_\mathrm{tunnel}>r_+-R$, so that the wave can enter and cancel the contribution of mode 2b, leading to no large asymmetry.
    }
\end{figure}

In order to understand the behavior for $(1-\delta)R<r\leq R$, let us re-express \eqref{schrodinger} in terms of $\varphi_\ell=r\psi_\ell$ as
\begin{gather}\label{eq:phi_schrodinger}
    -\frac1{2m}\partial_r^2\varphi_\ell+V_\mathrm{eff}(r)\varphi_\ell=\frac{k^2}{2m}\varphi_\ell,\\
    V_\mathrm{eff}(r)\equiv\frac{\ell(\ell+1)}{2mr^2}+U\cdot\Theta(R-r).
\label{eq:Veff}\end{gather}
\eqref{phi_schrodinger} has the exact form of the Schrodinger equation with an effective potential $V_\mathrm{eff}$.  Therefore, we can understand neutrino/antineutrino propagation near/inside the Earth in terms of an incoming wave scattering off this potential.  Several examples of $V_\mathrm{eff}$ are shown in \figref{potentials}.  The angular modes shown in \figref{potentials} correspond schematically to the rays shown in \figref{earthfig}.  For instance, the curve labeled 1a in \figref{potentials} shows the potential for an antineutrino mode with low enough angular momentum to penetrate deep into the Earth.  Classically, an incoming wave which is incident on this potential will be reflected at the ``turning point" $r_0\approx\ell/k'$ of the potential.  (In the ray optics picture, this corresponds to the impact parameter of the ray.)  The wave will therefore not contribute to the asymmetry at radii $r<r_0$.  The curve labeled 1b in \figref{potentials} corresponds to a neutrino ray with the same classical turning point.  Just as in \figref{earthfig}, the angular modes 1a and 1b contribute to the asymmetry at exactly the same set of points.  Thus, their contributions will cancel, leading to the small $\mathcal O(\delta)$ asymmetry derived above.

Consider instead the angular mode 2b with turning point $(1-\delta)R<r_0<R$.  The corresponding antineutrino mode 2a exhibits two turning points: $r_-\approx\ell/k'$ inside the Earth (at the same location as $r_0$ for mode 2b), and $r_+\approx\ell/k$ outside the Earth.  Classically, a wave incident on this potential will be reflected at $r_+$, and so will never reach the interior of the Earth.  In this way, the contribution from neutrino mode 2b inside the Earth is left uncanceled.  This results in the large asymmetry which can be observed for $(1-\delta)R<r\leq R$, in some cases.

Note that quantum mechanically, however, an incident wave may be able to tunnel through the external barrier of antineutrino mode 2a into the interior of the Earth.  The lower plot in \figref{potentials} shows in green a wave incident on the potential of mode 2a.  The wave begins to damp once it reaches $r_+$, with some damping length scale $L_\mathrm{tunnel}$.  If $L_\mathrm{tunnel}$ exceeds the distance $r_+-R$, then the wave will tunnel into the interior of the Earth without significant damping.  Consequently, it will be able to reach $r_-$ and cancel the contribution from neutrino mode 2b.  It is simple to estimate $L_\mathrm{tunnel}$ in the WKB approximation.  As the wave propagates from $r_+$ to $r_+-L$, the number of e-folds it damps is given by
\begin{align}\label{eq:WKB1}
    &\int_{r_+-L}^{r_+}\sqrt{2m\left(V_\mathrm{eff}(r)-\frac{k^2}{2m}\right)}dr\\
    &\quad\approx\int_{r_+-L}^{r_+}\sqrt{(r_+-r)\cdot\frac{2\ell(\ell+1)}
    {r_+^3}}dr\\
    &\quad=\frac23\sqrt{L^3\cdot\frac{2\ell(\ell+1)}{r_+^3}}.
\label{eq:WKB3}\end{align}
By setting \eqref{WKB3} equal to 1, we find that the tunneling range is roughly
\begin{equation}\label{eq:tunnel}
    L_\mathrm{tunnel}\sim\frac{r_+}{\ell^{2/3}}\sim\frac{(kR)^{1/3}}k.
\end{equation}
If we are interested in the asymmetry at the surface of the Earth, we should consider the mode with $r_-\approx R$, in which case $r_+-R\approx\delta R$.  Therefore, the asymmetry at the surface will be washed out by this tunneling effect when $L_\mathrm{tunnel}\gg\delta R$, or in other words, when $\delta^{3/2}kR\ll1$.  This justifies the behavior seen in \figref{contour}.

\section{Discussion}
\label{sec:discussion}

In this work, we have demonstrated that the large neutrino-antineutrino asymmetry derived in \citeR{arvanitaki}, under the assumption of a flat Earth, is only reproduced in the spherical calculation when \eqref{condition} is satisfied.  As the Earth is not large/dense enough to satisfy this condition, we showed that the asymmetry at the surface of the Earth is instead only $\mathcal O(\delta)$.  Importantly, the calculation of \secref{spherical} assumed that the Earth is perfectly spherical.  The physical values of the relevant length scales in this computation are $L_\mathrm{tunnel}\sim1\,\mathrm{m}$ and $\delta R\sim1\,\mathrm{cm}$.  The Earth is certainly far from a perfect sphere on these length scales, and so the presence of local terrain can affect the conclusions of this work.

We argue, however, that deviations from sphericity should typically make it more difficult to induce a large asymmetry.  To see why, let us return to the heuristic argument presented in \secref{heuristic}.  A large asymmetry occurs when there exists a large region underneath the surface of the Earth which is inaccessible to glancing antineutrino rays.  In particular, the extent of this region must be larger than the tunneling range $L_\mathrm{tunnel}$ derived in \eqrefRange{WKB1}{tunnel}.  Note that this derivation only relied on the effective potential outside the Earth, and so $L_\mathrm{tunnel}$ should not be affected by local deformations of the Earth.%
\footnote{The $R$ appearing in \eqref{tunnel} arises from the use of spherical coordinates, with the center of the Earth as our origin, rather than the presence of the Earth itself.}
The effect of local terrain should instead be to modify the extent of the inaccessible region.  The inaccessible region observed in the perfectly spherical case is a consequence of the smoothness of the Earth's surface, which limits the possible entry angles available to antineutrino rays.  As the surface becomes more inhomogeneous, a larger variety of entry angles become available, allowing the antineutrinos to access a larger region below the Earth's surface.  We therefore argue that the presence of local terrain will typically act to shrink the inaccesible region, implying that a large asymmetry will still be washed out by the tunneling argument presented in this work.

Before concluding, we contrast this work with \citeR[s]{huang,gruzinov,vantilburg2024wake}, which performed similar computations to the one outlined in \secref{spherical}.  As mentioned previously, \citeR{huang} computed the asymmetry at the surface of the Earth for the physical parameter values of the Earth ($kR=9\times10^9$ and $\delta=3\times10^{-8}$).  In order to make the computation numerically tractable, they utilized some unspecified Bessel function approximations, so we cannot directly check their result, but they do obtain an asymmetry $\Delta=-\delta$, and so their results agree with the findings of this work.  In contrast to \citeR{huang}, this work demonstrates that \eqref{condition} is necessary in order for the flat-Earth result to hold.

\citeR{gruzinov} utilizes a different approach to compute the neutrino-antineutrino asymmetry.  Rather than treating the C$\nu$B as monochromatic, they account for its full Fermi-Dirac distribution and instead appeal to thermal arguments to demonstrate that the asymmetry should be $\mathcal O(\delta)$.  In particular, their arguments imply that the asymmetry should be $\mathcal O(\delta)$, even when \eqref{condition} is satisfied. 
\figref{contour} clearly demonstrates that this is not the case for a monochromatic background, as most parameter values above the black line exhibit an asymmetry larger than $\mathcal O(\delta)$.  However, because there are some values which exhibit positive asymmetries and some which exhibit negative asymmetries, it is possible that after averaging over neutrino momentum, the total asymmetry is reduced to $\mathcal O(\delta)$.  These large negative asymmetries are simple to understand from \figref{earthfig}.  When \eqref{condition} is satisfied, ray 2a has a low probability to tunnel to the dashed red line, and so for most parameter values, we will observe an underdensity of antineutrinos.  However, for precisely the same reason that it is difficult for this ray to tunnel into the Earth, it is also difficult for it to tunnel \emph{out of} the Earth.  That is, if ray 2a does tunnel in, it will find itself in a long-lived bound state (sometimes called a ``whispering gallery mode"~\cite{whispering}), and so it can instead contribute a large antineutrino \emph{overdensity}.  As can be seen in \figref{contour}, the negative-asymmetry resonances corresponding to these bound states can be quite narrow, and so integrating over neutrino momentum precisely requires very good resolution in momentum.  We therefore do not attempt to verify the claims of \citeR{gruzinov}, but instead simply present \eqref{condition} as a different argument for why the asymmetry should be small.  We do note though that the arguments in \citeR{gruzinov} do not rely on the particular geometry of the Earth, and so apply even when the asphericity of the Earth is accounted for.

Finally, Appendix C of \citeR{vantilburg2024wake} also performs a number of numerical computations related to the problem considered in this work.  \citeR{vantilburg2024wake} simulates scattering of a thermal background, in both 2+1 and 3+1 dimensions.  They find $\mathcal O(\delta)$ effects which are consistent with the thermal arguments advanced in \citeR{gruzinov}.

\acknowledgments

We thank Asimina Arvanitaki for numerous fruitful discussions related to this work, including discussions regarding the condition in \eqref{condition}.  We also thank Savas Dimopoulos, Marios Galanis, Guo-yuan Huang, Zhen Liu, and Ken Van Tilburg for other discussions on this topic.  S.K. is supported by the U.S. Department of Energy, Office of Science, National Quantum Information Science Research Centers, Superconducting Quantum Materials and Systems Center (SQMS) under contract number DE-AC02-07CH11359.

\bibliographystyle{JHEP}
\bibliography{references.bib}

\providecommand{\href}[2]{#2}\begingroup\justify\begin{thebibliography}{10}

\bibitem{kolb}
E.~Kolb and M.~Turner, \emph{The Early Universe}.
\newblock Westview Press, 1994.

\bibitem{baumann}
D.~Baumann, \emph{Cosmology}.
\newblock Cambridge University Press, 2022.

\bibitem{yanagisawa}
C.~Yanagisawa, \emph{Looking for cosmic neutrino background}, \href{https://dx.doi.org/10.3389/fphy.2014.00030}{\emph{Frontiers in Physics} {\bf 2} (2014) }.

\bibitem{scott}
D.~Scott, \emph{The cosmic neutrino background},  \href{https://arxiv.org/abs/2402.16243}{{\tt arXiv:2402.16243}}.

\bibitem{weinberg}
S.~Weinberg, \emph{Universal neutrino degeneracy}, \href{https://dx.doi.org/10.1103/PhysRev.128.1457}{\emph{Phys. Rev.} {\bf 128} (Nov 1962) 1457--1473}.

\bibitem{ptolemy}
E.~Baracchini, M.~G. Betti, M.~Biasotti, A.~Bosca, F.~Calle, J.~Carabe-Lopez et~al., \emph{Ptolemy: A proposal for thermal relic detection of massive neutrinos and directional detection of mev dark matter},  \href{https://arxiv.org/abs/1808.01892}{{\tt arXiv:1808.01892}}.

\bibitem{stodolsky}
L.~Stodolsky, \emph{Speculations on detection of the "neutrino sea"}, \href{https://dx.doi.org/10.1103/PhysRevLett.34.110}{\emph{Phys. Rev. Lett.} {\bf 34} (Jan 1975) 110--112}.

\bibitem{arvanitaki}
A.~Arvanitaki and S.~Dimopoulos, \emph{Cosmic neutrino background on the surface of the earth}, \href{https://dx.doi.org/10.1103/PhysRevD.108.043517}{\emph{Phys. Rev. D} {\bf 108} (Aug 2023) 043517}.

\bibitem{huang}
G.~Huang, \emph{Neutrino-antineutrino asymmetry of c$\nu$b on the surface of the round earth},  \href{https://arxiv.org/abs/2401.07347}{{\tt arXiv:2401.07347}}.

\bibitem{gruzinov}
A.~Gruzinov and M.~Mirbabayi, \emph{The density of relic neutrinos near the surface of earth},  \href{https://arxiv.org/abs/2403.03152}{{\tt arXiv:2403.03152}}.

\bibitem{vantilburg2024wake}
K.~V. Tilburg, \emph{Wake forces},  \href{https://arxiv.org/abs/2401.08745}{{\tt arXiv:2401.08745}}.

\bibitem{github}
\url{https://github.com/skalia618/CnuB-Asymmetry}.

\bibitem{whispering}
\url{https://en.wikipedia.org/wiki/Whispering-gallery_wave}.

\end{thebibliography}\endgroup

\end{document}